\documentclass[aps,twocolumn,amsmath,amssymb]{revtex4}
\usepackage{graphicx}
\usepackage{dcolumn}
\usepackage{bm}
\usepackage{amssymb}
\usepackage{amsmath}
\usepackage{epsf}

\begin{document}

\title{Connectivity and Dynamics of Neuronal Networks as Defined 
by the Shape of Individual Neurons}

\author{Sebastian Ahnert}
\affiliation{Theory of Condensed Matter,
Cavendish Laboratory,
University of Cambridge,
JJ Thomson Avenue,
Cambridge CB3 0HE, UK}
\author{Luciano da Fontoura Costa}
\email{luciano@if.sc.usp.br}
\affiliation{Instituto de F\'{\i}sica de S\~{a}o Carlos,
Universidade de S\~{a}o Paulo, Av. Trabalhador S\~{a}o Carlense 400,
Caixa Postal 369, CEP 13560-970, S\~{a}o Carlos, S\~ao Paulo, Brazil \\}

\date{7 September 2007}

\begin{abstract}
Neuronal networks constitute a special class of dynamical systems, as
they are formed by individual geometrical components, namely the
neurons.  In the existing literature, relatively little attention has
been given to the influence of neuron shape on the overall
connectivity and dynamics of the emerging networks.  The current work
addresses this issue by considering simplified neuronal shapes
consisting of circular regions (soma/axons) with spokes
(dendrites). Networks are grown by placing these patterns randomly in
the 2D plane and establishing connections whenever a piece of dendrite
falls inside an axon.  Several topological and dynamical properties of
the resulting graph are measured, including the degree distribution,
clustering coefficients, symmetry of connections, size of the largest
connected component, as well as three hierarchical measurements of the
local topology.  By varying the number of processes of the individual
basic patterns, we can quantify relationships between the individual
neuronal shape and the topological and dynamical features of the
networks. Integrate-and-fire dynamics on these networks is also
investigated with respect to transient activation from a source node,
indicating that long-range connections play an important role in the
propagation of avalanches.
\end{abstract}


\maketitle

\vspace{0.5cm}
\emph{`Nothing in excess.' (Delphic proverb)}

\section{Introduction}

Many discrete systems in nature can be modeled in terms of networks
which explains the remarkable development of the field of complex
networks research over the last decade (e.g.~\cite{Albert_Barab:2002,
Dorogov_Mendes:2002, Newman:2003, Boccaletti:2006, Costa_surv:2007}).
Because such systems involve a large number of elements exchanging
mass, energy or information, their representation as a graph or
network is intrinsic.  A wide range of complex systems, from the
Internet to protein-protein interaction, has been successfully mapped
and studied in terms of complex networks
(e.g.~\cite{Costa_appls:2008}).  The connectivity of a network also
influences the properties of dynamical processes which may take place
on it. This relationship between topology and dynamics is of
particular interest in many networks \cite{Costa_revneur:2005}
including neuronal ones.

Despite the many applications of complex networks research, little
attention has been given to systems in which the overall connectivity
is affected (or even defined) by the geometrical features of the
individual constituent elements.  Indeed, the majority of works
addressing structure and dynamics in complex networks tends to equate
the structure with the topology of the networks, without regarding the
geometry of the involved components.  While it is true that many
networks do not have a geometrical basis --- such as author
collaboration and disease transmission networks, as well as the
Internet --- some important complex systems are formed by elements
with a well-defined geometry which influences or even defines the
emerging connectivity and dynamics.  Such \emph{morphological
networks} include protein-protein interaction as well as the
biological neuronal networks in the central nervous systems (CNS) of
animals, which are called \emph{morphological neuronal networks}.  In
such networks which are the result of spatial interactions between
geometrical constituent elements, the information about the geometry
of the original elements may not be available to the researcher.  This
is often the case in protein-protein interaction and neuronal
networks, where only the connectivity between the nodes is provided.

A particularly important example of morphological networks are
neuronal systems, which can be represented as directed networks with
each neuron being a node and each synapse a directed edge.  The
resulting connectivity and dynamics are closely related to the
geometry of the involved neurons.  Indeed, a large number of
morphological neuronal types (e.g.~\cite{Squire:2003}) have been
identified in the CNS of vertebrates as well as more primitive animals
(e.g. insects).  Morphological neuronal types range from cells as
simple as the bipolar neurons of the retina to the intricate Purkinje
neurons of the cerebellum.  Because the neuronal shapes are highly
variable, no consensus exists on how they should be categorized.

The contribution of the individual morphology of neurons to the
overall network connectivity mostly occurs in two ways: (a) the
biochemical differentiation of neuronal types constrains the
respective geometries and connections (e.g. a Purkinje cell will never
exhibit radial organization); and (b) the growth of each dendrite or
axon takes into account influences from the surrounding environment,
such as electric fields and gradients of concentration of molecules.
Therefore, the shape of a neuron in the CNS is a consequence of both
\emph{nature} (i.e. the biochemical content associated to cell
differentiation) and \emph{nurture} (i.e. the effects of the
surrounding environment).  The spatial distribution of the bodies of
the neurons (called somata or perykaria) also plays an important role
in defining the overall connectivity.  It is important to bear in mind
that neuronal systems are highly plastic, undergoing structural and
dynamical changes during their whole lifetime.  Therefore, it is
particularly important to consider the interrelationship between
neuronal structure and dynamics in growing networks.

Here we provide the first systematic investigation on the relationship
between the geometry and dynamics of growing morphological networks.
Because of the stochastic nature of biological neuronal networks, it
becomes essential to consider a large number of realizations in order
to obtain meaningful measurements of topology and dynamics.
Therefore, we keep the total number of involved parameters as small as
possible, focussing on those which are clearly related to neuronal
shape and connectivity.  First, we restrict our investigation to 2D
neurons and neuronal networks.  Such a simplification is justified by
the fact that several real neurons such as retinal ganglion cells,
Purkinje and even the basal dendritic structure of the ubiquitous
pyramidal cells are mainly planar.  Next, we adopt prototypical radial
neurons involving a central region (axon) from which a given number of
dendrites with the same length (spokes) emerge.  This model implies as
parameters: the number and length of dendrites, angular distribution
of the dendrites, and the radius of the central region.  By assuming a
fixed dendritic area for all neurons, the length of the dendrites can
be related to their number, so that one of these parameters can be
omitted.

Once the types of neurons are defined, the networks can be obtained by
progressively placing new neuronal cells and establishing
connections whenever a part of a dendrite touches an axon.  We use a 
uniform random distribution of the position of the cells.  Before
being added to the network, each cell is rotated by a uniformly random
angle.  Therefore, the only additional parameters implied by the
network growth are the current number of added cells and the size of
the space along which the neurons are distributed.

In order to investigate the relationship between structure and
dynamics in these networks, we make a series of different
measurements. In order to express the topology of the networks we
determine the degrees of the nodes, their clustering coefficient,
Garlaschelli's symmetry index~\cite{Garlaschelli:2004}, as well as the
size of the largest connected component in the network.  The dynamics
of the respective networks is obtained in terms of integrate and fire
dynamics~\cite{Koch:1998,Koch_Segev:1998}.  More specifically, each
node is understood as a neuron which integrates the received
activation until a threshold is reached, in which case a spike is
produced as output.  It has been
verified~\cite{Costa_equiv:2008,Costa_HAL:2008} that
integrate-and-fire complex networks can undergo avalanches of
activations when stimulated from individual nodes, with the type of
connectivity substantially affecting the dynamics.  More specifically,
such avalanches have been related to specific concentric organization
of the network connectivity~\cite{Costa_equiv:2008,Costa_HAL:2008}.
Given the source node $i$, the nodes which are at topological distance
1 from $i$ are called the first concentric level, the neurons which
are at distance 2 constitute the second hierarchical level, and so on.
Avalanches are related to the existence of concentric levels with a
large number of nodes.  More specifically, if a concentric level $h$
contains many nodes while the preceding and subsequent levels are less
populated, the firing of neurons in the level $h$ tends to induce
overall activation of the neighboring levels, which then propagates
through the whole network.  In the case of undirected networks it is
possible to obtain a simple equivalent model of the original network,
formed by a chain of a few equivalent nodes across the hierarchical
levels~\cite{Costa_equiv:2008,Costa_HAL:2008}. This model allows
predictions of avalanche characteristics, such as the required
activation ratios~\cite{Costa_HAL:2008}, even for symmetrized versions
of directed networks.  Such an approximation is particularly justified
when the original directed network involves a relatively high degree
of symmetric connections, as is the case for our morphological
networks with large number of spokes $n$.

Possible interdependencies between the structure and dynamics
of the networks are investigated by considering the
Pearson correlation of these measurements.

We begin with a brief overview of the basic concepts of complex
networks theory and morphological neuronal networks.

\section{Basic Concepts}

\subsection{Complex Networks Topology}

Directed networks are defined by a set of $N$ nodes and $E$ edges
which can be represented by an adjacency matrix $A$, such that its
element $a_{ij}=1$ indicates the presence of a directed edge from node
$i$ to node $j$, while $a_{ij}=0$ expresses its absence.

The topological properties of a network can be quantified and
characterized in terms of a comprehensive set of
measurements~\cite{Costa_surv:2007}.  In this work we consider the
following features: (i) in- and out-degree; (ii) clustering
coefficient; (iii) Garlaschelli's symmetry index; and (iv) size of the
largest connected component in the network.  More specifically, we
consider the averages and standard deviations of the measurements
taken over the whole networks.  Each of these measurements, as well as
the motivation for their adoption, are described in detail as follows:

\emph{Average Node In- and Out-degree:} The in- and out-degree of an
individual node $i$ corresponds to the number of edges leading to and
from that node.  These measurements can be obtained directly from the
adjacency matrix as:

\begin{eqnarray}
  a_i^{in} = \sum_{j=1}^{N} a_{ji} \\
  a_i^{out} = \sum_{j=1}^{N} a_{ij} 
\end{eqnarray}

The average in- and out-degrees, $\left< k_i^{in}
\right>$ and $\left< k_i^{out} \right>$, taken over the whole network, 
are identical and provide a quantification of the overall degree of
connectivity in the network.

\emph{Average Clustering Coefficient:} The clustering coefficient of a
node $i$ measures how interconnected the neighbors of node $i$ are. It
does so by counting the number of connections between neighbors of $i$
and dividing it by the number of possible connections - in other
words, by the number of pairs of neighbors. In directed networks this
can be written as:

\begin{equation}
c_i = \frac{n}{d_i(d_i-1)}
\end{equation}

where $d_i = \sum_j (a_{ij} + a_{ji})$ is the number of neighbors of
node $i$ and $n = \sum_{j,k } [1 - (1 - a_{ij})(1 - a_{ji})][1 - (1 -
a_{ik})(1 - a_{ki})](a_{jk} + a_{kj})$ is the number of directed edges
between those nodes.  Note that $0 \leq c_i
\leq 1$, with $c_i=0$ indicating a total lack of interconnectivity 
between the neighbors and $c_i=1$, i.e. all neighbors are connected to
each other. Another version of the clustering coefficient for directed
networks, which takes the variety of directed topological environments
into account, has been reported in ~\cite{Fagiolo}.

\emph{Garlaschelli's Symmetry Index:} Introduced by
Garlaschelli~\cite{Garlaschelli:2004}, this measurement quantifies the
symmetry of connections between pairs of nodes relative to an
Erd\"os-R\'enyi network of the same size and density.  Let $E$ be the
total number of directed edges in the network, and $E_b$ be the total
number of bidirectional (symmetric) edges, and define the ratios $r =
E_b / E$ and $a = E/(N^2-N)$.  The symmetry index is then given by:

\begin{equation}
  \rho = \frac{r-a}{1-a}  \label{eq:rho}
\end{equation}

\emph{Size of the Largest Connected Component:} At any time during the
growth of a network, it is of interest to quantify the overall path
connectivity among the existing nodes.  This can be done by measuring
the size $C$ of the largest connected component in the network.  More
specifically, a \emph{connected component}~\footnote{It is important
to observe that, in the case of directed graphs, the terminology
\emph{strongly connected components} should be preferred.} is a
subgraph such that each node can be reached from any other node
through at least one path. 

Because the dynamics of complex neuronal networks is defined by the
concentric organization
(e.g.~\cite{Costa_equiv:2008,Costa_eqcomm:2008}) of their topology, we
also consider three hierarchical measurements, namely the hierarchical
number of nodes, the hierarchical degree and the intra-ring degree.
Given an undirected network, the concentric level $h$ of a node $i$ is
defined as the set of nodes which are separated from $i$ by a shortest
path of length $h$. The maximum value of $h$ is $H$. The
\emph{hierarchical number of nodes} at level $h$, with respect to a
reference $i$, is equal to the number of nodes at that level. The
\emph{hierarchical degree} of a node $i$ at level $h$ corresponds to
the number of edges between the levels $h$ and $h+1$.  The
\emph{intra-ring degree} of a node $i$ at level $h$ is the number of
edges between nodes of that level.

\subsection{Morphological Networks}

Morphological networks are networks formed by a spatial distribution
of individual geometrical components~\cite{Costa_Manoel:2003} with
connections between them defined by their overlap.  For instance, a
biological neuronal network is composed of neurons with a specific
geometry given by their dendritic and axonal arborizations.
Morphological networks can be classified into subcategories with
respect to: (i) the individual shape of the basic elements, (ii) the
degree of homogeneity of such elements in the network (e.g. networks
containing a single or multiple type of individual shape), and (iii)
the way in which these elements are spatially distributed.  Each of
these cases is discussed below in detail:

\emph{Individual Neurons:} The individual shape of the network elements
can closely mirror real elements (e.g. by using images of neurons) or
they can take a more abstract form (e.g. by using some pattern
generation method).  In the case of neural morphological networks, the
shape of the neurons is composed of dendrites and axons.  As neuronal
connections (i.e. synapses) extend from axon to dendrite, such
networks are directed.  The shape of the elements can be represented
in a discrete form, such as digital images or a continuous form,
e.g. by (possibly piecewise) continuous curves.  The geometrical
properties of the basic elements in the network can be quantified by
using a series of measurements such as area, perimeter, fractal
dimension, symmetry, etc.

\emph{Degree of Homogeneity:} Morphological networks may involve single
or multiple basic elements, possibly involving geometrical
transformations (e.g. rotation or scaling).

\emph{Spatial Distribution:} Morphological networks are created by 
distributing the element shapes in space and considering their
overlap. Several types of distributions can be considered,
e.g. uniformly random or preferential to specific positions (e.g. a
normal distribution centered at a given point).

In order to keep the number of involved parameters and degrees of
freedom as small as possible, allowing statistically representative
simulations, we choose the following representation of individual
neurons: (a) the dendrites are represented by a star of spokes with
uniform angle between them; (b) the total length of the dendrites is
fixed and therefore independent of the number of spokes; (c) the axon,
which corresponds to the soma, is a circle of fixed radius at the
center of the star.  Identical neurons, rotated randomly, are used to
build each network.  The neurons are placed sequentially at uniformly
random positions within the unit square.

\begin{figure}[htb]
  \begin{center} 
  \includegraphics[width=1\linewidth,angle=0]{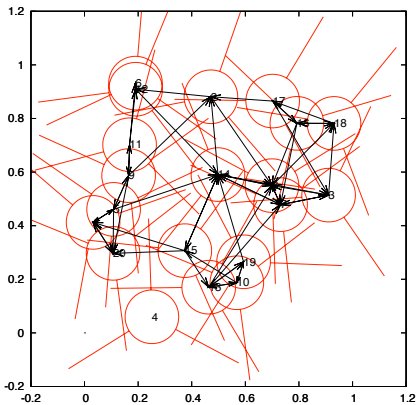}
  \caption{An example of morphological network obtained by considering 
              $N=20$ and $n=3$.  A directed 
              connection is established whenever
              a dendrite crosses the axon of another neuron.
  ~\label{fig:ex_net}} 
  \end{center}
\end{figure}

Figure~\ref{fig:ex_net} illustrates a simple morphological network
obtained for $N=20$ and $n=3$.  Each neuron corresponds to a circle
(the soma and axon) from which $n=3$ spokes emanate.  The neurons,
which are placed with varying rotations, are identified by their
respective central points.  The network obtained is shown in black,
superimposed into the original morphological structure.  Observe that
neuron number 4 remains isolated.  All other 19 cells belong the
largest strongly connected component.
  
\subsection{Integrate-and-Fire Complex Neuronal Networks}

\begin{figure}[htb]
  \begin{center}
  \includegraphics[width=1\linewidth]{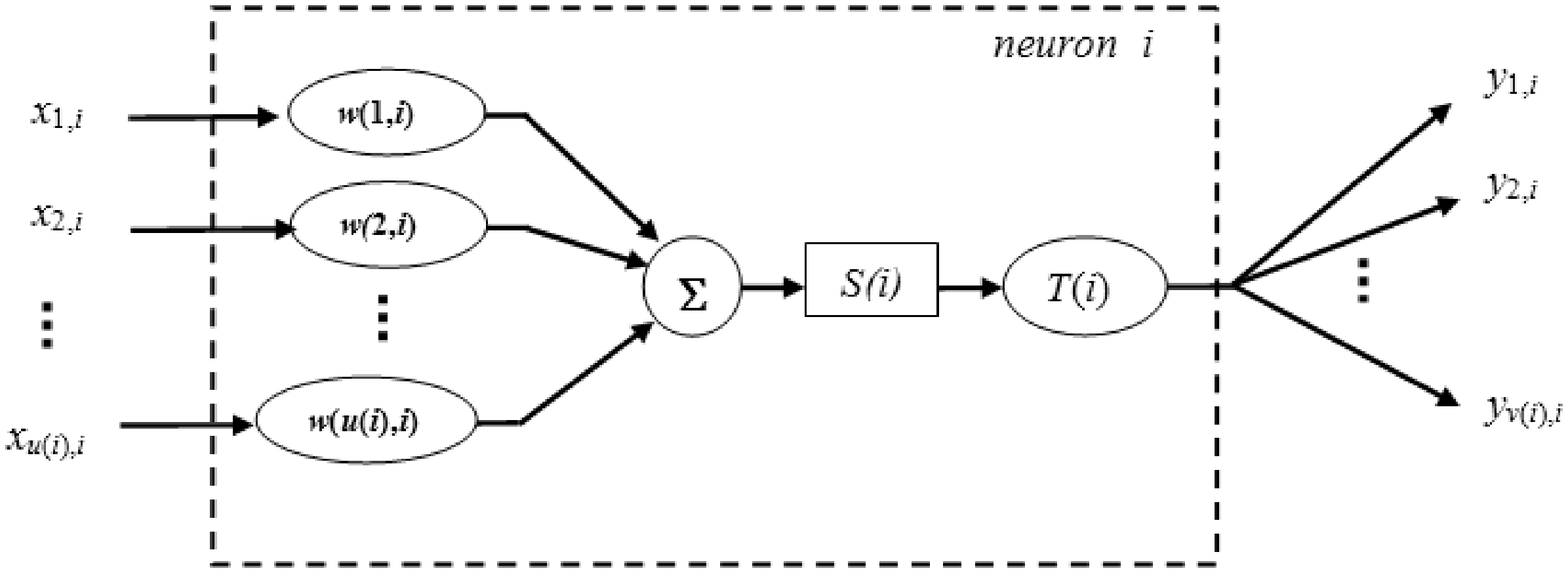} 
  \caption{The integrate-and-fire model of neuronal cell dynamics
             adopted in this work incorporates $v(i)$ incoming
             connections with respective weights, $u(i)$ outgoing 
             links, an integrator $\Sigma$, the internal memory
             $S(i)$ storing the time-integrated activation, and the
             non-linear transfer function (a hard limit is adopted
             in this work).  Every time the internal activation $S(i)$
             exceeds the threshold $T(i)$, the neuron fires, generating
             a spike of fixed intensity $1$ and clearing the internal
             activation.
  ~\label{fig:neuron}}
  \end{center}
\end{figure}

In this work, the dynamics of the morphological networks is simulated
by considering the integrate-and-fire neuronal model
(e.g.~\cite{Koch:1998}), which incorporates the most important
elements characterizing real neuronal networks, namely integration of
the activation received by each cell along time, related to the
biological phenomenon of \emph{facilitation}~\cite{Squire:2003}, and a
non-linear transfer function. Figure~\ref{fig:neuron} illustrates the
main components of the integrate-and-fire neuronal cell.  The incoming
axons are connected to the dendrites of the neuronal cell $i$, their
signals being added by the integrator $\Sigma$ at each time step and
stored in the internal state $S(i)$.  In this work we assume equal
synaptic weights.  Once the value of $S(i)$ reaches a given threshold
$T(i)$, the cell produces a spike and its internal state $S(i)$ is
cleared.  The activation conveyed by the axons is fixed at intensity
1, reflecting the constant amplitude of the action potential
characteristic of real neuronal networks.

Though other types of activations can be considered, in the present
work we activate the networks through a \emph{source node}.  More
specifically, one of the network nodes is chosen and injected with
constant activation equal to $1$.  All neurons have the same threshold
$T(i)=T$, and all synaptic weights are fixed at $1$.

The networks studied here are of sizes $N=200$ and $400$, with a
varying number of dendrites (i.e. spokes).  The radius of the soma was
0.0625 and the total length of the dendrites was 0.5. Only the largest
connected components are considered for the analysis of the
integrate-and-fire dynamics.

The overall integrate-and-fire dynamics unfolding in a network can be
characterized in several ways (e.g.~\cite{Koch:1998, Costa_nrn:2008,
Costa_equiv:2008}).  Here, we consider the internal activation and
spikes produced by each neuron over time.  More specifically, we will
focus attention on the transient phenomenon of
\emph{avalanches of spikes} (e.g.~\cite{Costa_nrn:2008, Costa_equiv:2008, 
Kaiser:2007}) which may take place during the initial activation of
the network.  Because we are considering the largest connected
component in each network, the activation arriving at the source
eventually reaches all the neurons in the networks.  Interestingly,
such an activation may progress either gradually or involve an abrupt
onset of spikes, i.e. an avalanche.  The intensity and sharpness of
these avalanches are defined by the concentric properties of the
topology of the respective networks~\cite{Costa_nrn:2008,
Costa_equiv:2008}.  After an avalanche, the total number of spikes in
the network tends to exhibit well-defined
oscillations~\cite{Costa_nrn:2008}.

The type of activation transition can be
predicted~\cite{Costa_lattice:2008} by considering the three following
ratios derived from the hierarchical number of nodes $n_h(i)$,
hierarchical degrees $k_h(i)$, and intra-ring degrees $a_h(i)$

\begin{eqnarray}
  s1_h(i) = \frac{k_h(i)}{n_{h+1}(i)} \frac{1}{T} \label{eq:s1} \\
  s2_h(i) = \frac{a_h(i)}{n_{h}(i)} \frac{1}{T} \label{eq:s2} \\
  s3_h(i) = \frac{k_h(i-1)}{ n_{h-1}(i)} \frac{1}{T} \label{eq:s3} \\
\end{eqnarray}

The quantity $s1_h(i)$, called \emph{forward activation ratio},
quantifies the intensity of the transfer of activation from the
neurons in level $h$ into level $h+1$ after most of the neurons in
level $h$ have fired (because of degree regularity, the nodes within
each concentric level tend to fire at similar times).  The
\emph{reflexive activation ratio} $s2_h(i)$ expresses how much of the
activation of the neurons at level $h$ remains at that same level.
The transfer of activation from level $h$ to level $h-1$ is quantified
in terms of the
\emph{backward activation ratio} $s3_h(i)$.  All ratios are normalized 
with respect to the threshold $T$ of every cell so that ratio values
of $1$ indicate a full transfer of the received activation.

It is important to note that the propagation of the avalanches is
influenced by the finite size and local topology of the networks.
Consider a perfectly regular but finite orthogonal lattice.  For any
given node, the concentric layers of neighbours surrounding it will be
characterized by levels with progressively increasing number of nodes,
up to a point where this number starts to decline again as a
consequence of the finite size of the network. This maximum
(see~\cite{Costa_HAL:2008} for the complete analytical results), plays
an important role in the formation of avalanches.  Several other types
of complex networks have been shown to undergo avalanches under
specific parameter
configurations~\cite{Costa_equiv:2008,Costa_HAL:2008}. It is also
possible to design networks which will never produce avalanches,
e.g. by having more uniform distributions of nodes along the
concentric levels.  Real-world neuronal systems are often organized
into modules with sometimes only a few dozens of neurons, which is in
line with the size of networks in our models.

\section{Characterization of the Topology}

\begin{figure*}[htb]
  \begin{center} 
  \includegraphics[width=0.9\linewidth]{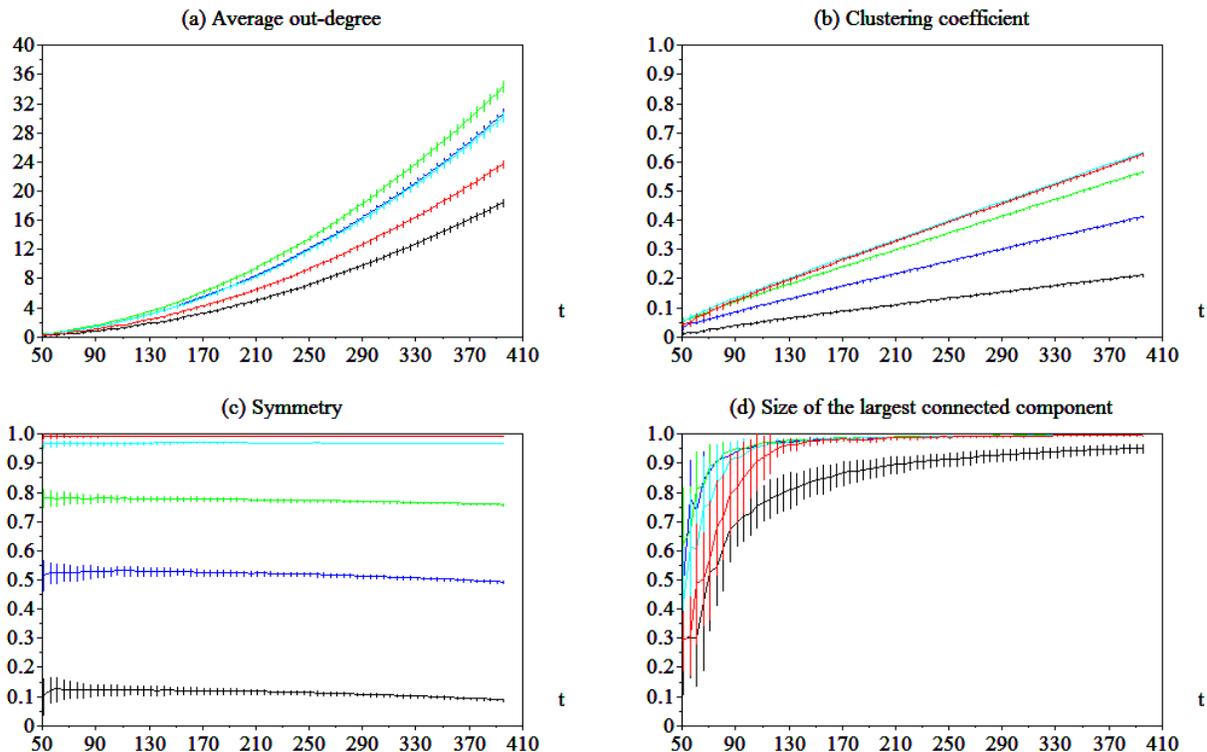}
  \caption{The average out-degree $\left< ok \right>$ (a), average
  clustering coefficient $\left< cc \right>$ (b), symmetry coefficient
  $\rho$ (c) as well as the size of the largest connected component
  divided by the number of neurons (d) in terms of the time obtained
  along the growth of 50 morphological networks with $n = 1$ to $20$
  spokes ($n=1$ in black, $n=3$ in blue, $n=5$ in green, $n=10$ in
  cyan, $n=20$ in red).~\label{fig:meas_200}} 
  \end{center}
\end{figure*}

Figure~\ref{fig:meas_200} shows the average out-degree $\left<
k^{out}_i \right>$ (a) , the average clustering coefficient $\left<
c_i \right>$ (b), the symmetry coefficient $\rho$ (c) as well as the
size of the largest connected component divided by the number of
neurons (d) in terms of the time $t$ (which is equal to the number of
placed neurons) for 50 morphological networks with $n = 1$ to $20$
spokes (identified by different colors).

It is clear from Figure~\ref{fig:meas_200}(a) that the average
in-degree increases roughly quadratically as additional neurons are
incorporated into the network. In addition, the average in-degree also
increases with the number of spokes up to $n=5$, decreasing
thereafter.  As shown in Figure~\ref{fig:meas_200}(b), the average
clustering coefficient tends to increase linearly with the number of
added neurons $t$.  It also increases with the number of spokes $n$ up
to $n=10$, before leveling off at higher values of $n$.  The symmetry
index, shown in Figure~\ref{fig:meas_200}(c), remains nearly constant
with $t$ but increases steadily with the number of spokes $n$.  This
reflects the fact that for large numbers - and thus a high angular
density - of spokes, there is a high probability that if the axon of
neuron A is reached by the dendrite of neuron B, the opposite will
also be true.  The size of the largest connected component $C$ always
converges to 1, with faster convergence being observed for larger
values of $n$.  Except for $n=1$, most of the neurons will be part of
the largest connected component for $t >100$.

\begin{figure*}[htb]
  \begin{center} 
  \includegraphics[width=1\linewidth]{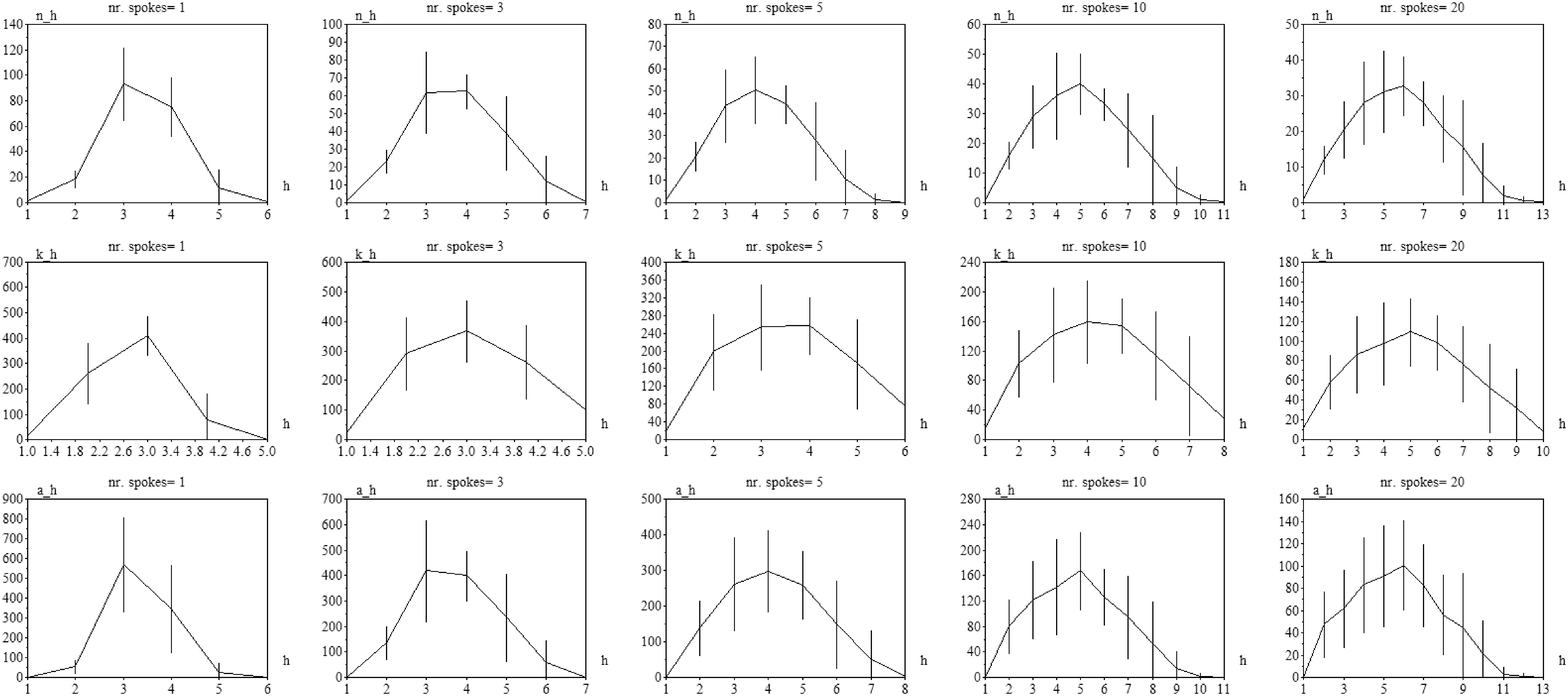}
  \caption{The hierarchical number of nodes $n_h(i)$, the hierarchical 
             degrees $k_h(i)$ and the intra-ring degrees $a_h(i)$
             obtained for the morphological networks with $N=200$
             and $n=1, 3, 5, 10$ and $20$ spokes.
  ~\label{fig:conc_meas}} 
  \end{center}
\end{figure*}

The concentric organization of the adopted morphological networks
($N=200$), known to play an important role in the integrate-and-fire
dynamics, is illustrated in Figure~\ref{fig:conc_meas} with respect to
the average standard deviation of the hierarchical number of nodes
$n_h(i)$, hierarchical degrees $k_h(i)$, and intra-ring degrees
$a_h(i)$.

Figure~\ref{fig:meas_200} makes it clear that all three measurements
behave similarly, exhibiting a peak near the middle concentric level.
In addition, larger number of spokes tend to increase the total number
$H$ of concentric levels in the networks, which goes from about 6 for
$n=1$ to 13 for $n=20$.  Similar results were obtained for $N=400$.

\begin{figure*}[htb]
  \begin{center} 
  \includegraphics[width=0.3\linewidth]{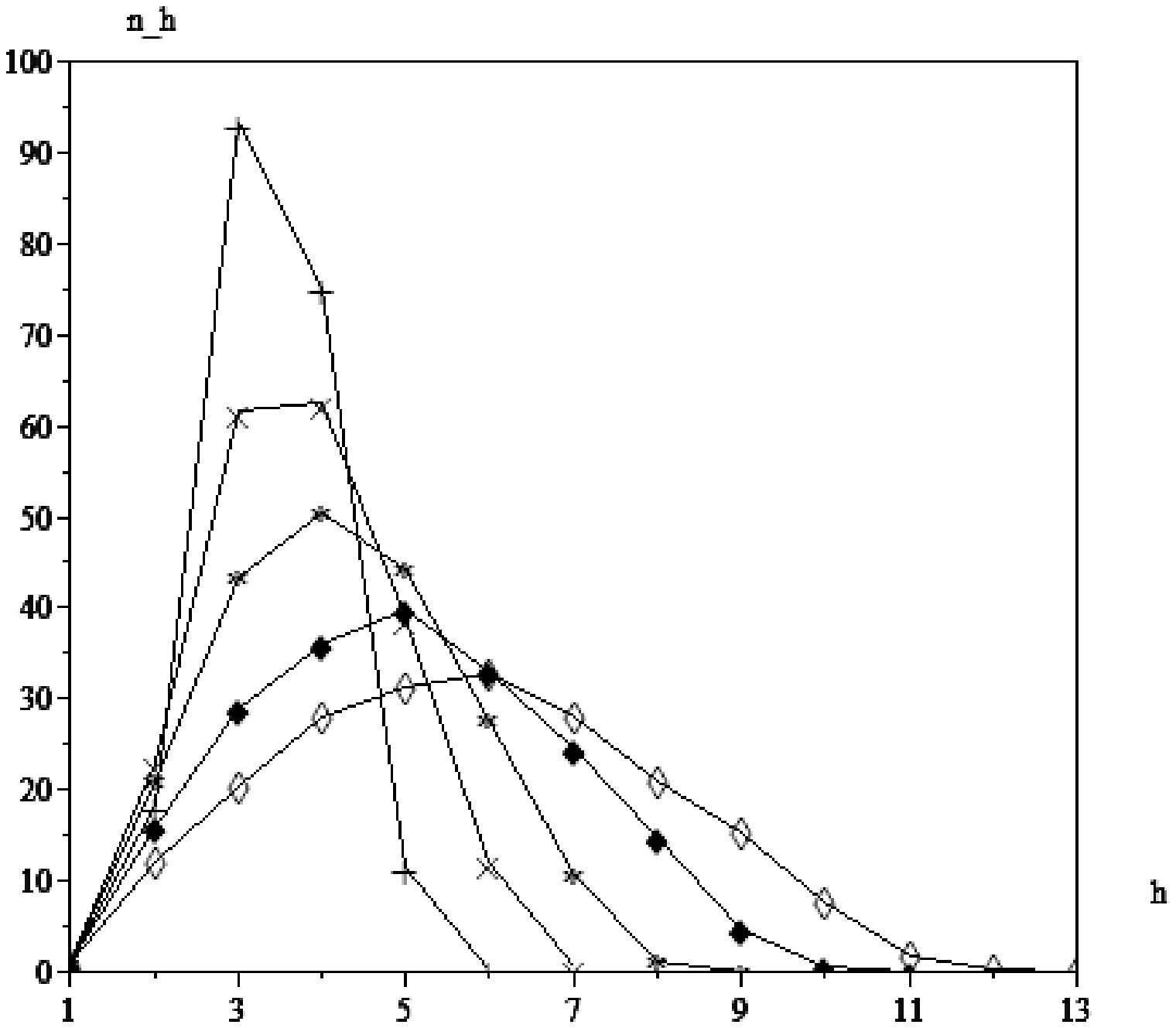}
  \includegraphics[width=0.3\linewidth]{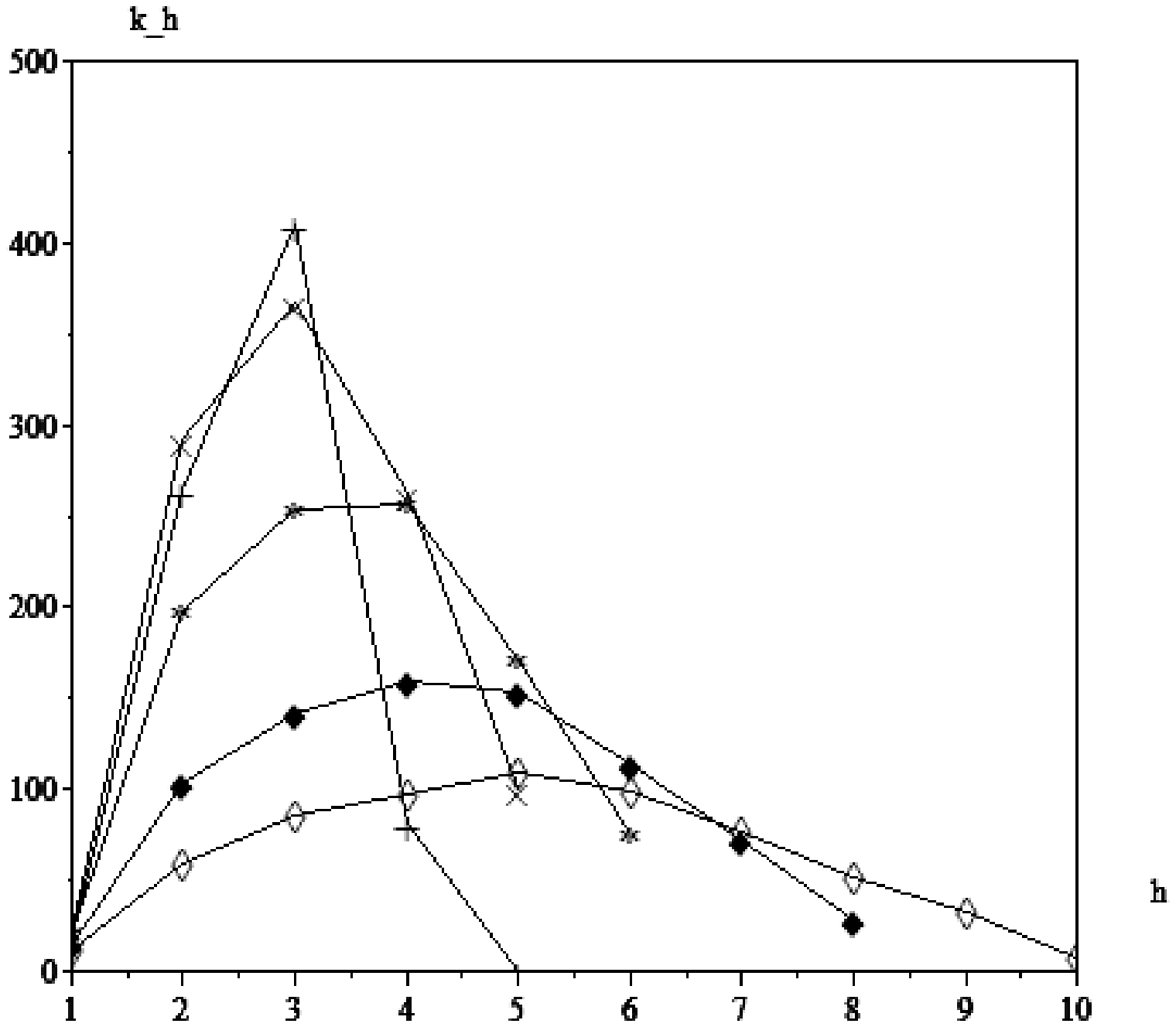}
  \includegraphics[width=0.3\linewidth]{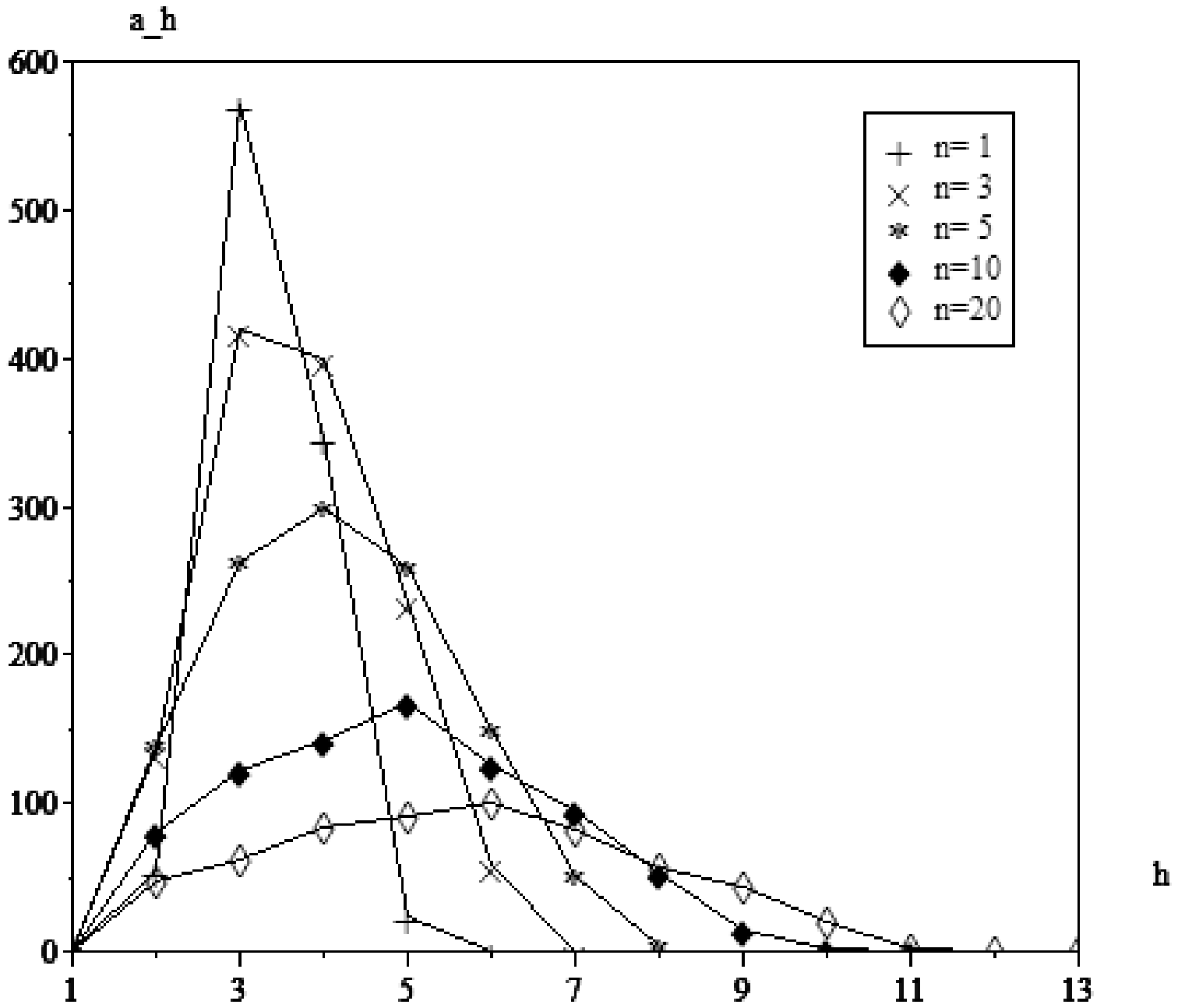} \\
  (a) \hspace{5cm}   (b)  \hspace{5cm}   (c)
  \caption{Superposition of the mean values of the three considered
            concentric measurements for $n=1, 3, 5, 10$ and $20$.
  ~\label{fig:PCA}} 
  \end{center}
\end{figure*}

Figure~\ref{fig:PCA} shows the mean values of the three considered
concentric measurements for $n=1, 3, 5, 10$ and $20$.  It is clear
from this figure that the increase of the number of spokes implies a
larger number of concentric levels as well as smoother distributions
of measurements along the $h-$axis.  Furthermore it is evident that
the dispersions of each configuration decrease with $n$.  Because the
integrate-and-fire dynamics is defined by the concentric organization
of the network, the distribution of the configurations shown in
Figure~\ref{fig:PCA} implies that the adopted networks are poised to
yield distinct types of avalanche dynamics.

In order to obtain additional insights about the latter issue, namely
the type of activation transitions during the transient regime, we
calculated also the three activation ratios
(Eqs.~\ref{eq:s1}--~\ref{eq:s2}) for each of the configurations of
morphological networks.  Table~\ref{tab:s1s2s3_short} shows the mean
and standard deviation values of the three activation ratios obtained
for the two concentric levels associated to the largest number of
original nodes assuming $T=7$.  It should be recalled that these three
ratios are defined for undirected networks, so that the morphological
structures were symmetrized before the respective calculations.  It
has been verified that the ratios obtained for the symmetrized
versions of a directed network still capture to a good deal the
integrate-and-fire important dynamical features.

\begin{table*}
\centering
\begin{tabular}{|c||c|c|c|c|c|c|}  \hline  
$n$    &  $s1 (first)$   & $s1 (second)$     & $s2 (first)$    &  $s2 (second)$         & $s3 (first)$    & $s3 (second)$    \\  \hline
$n=1 $ &2.27$\pm$0.43&1.57$\pm$0.53 &2.01$\pm$0.22&1.24$\pm$0.41 &4.27$\pm$1.28&2.08$\pm$1.66 \\ \hline
$n=3 $ &2.46$\pm$0.48&2.42$\pm$0.66 &2.42$\pm$0.24&2.04$\pm$0.27 &3.48$\pm$1.32&2.38$\pm$0.96 \\ \hline
$n=5 $ &2.16$\pm$0.42&1.96$\pm$0.42 &2.17$\pm$0.28&1.95$\pm$0.28 &2.63$\pm$0.75&2.21$\pm$0.82 \\ \hline
$n=10$ &1.71$\pm$0.36&1.55$\pm$0.25 &1.52$\pm$0.22&1.43$\pm$0.20 &1.91$\pm$0.41&1.81$\pm$0.53 \\ \hline
$n=20$ &1.30$\pm$0.26&1.29$\pm$0.30 &1.17$\pm$0.19&1.05$\pm$0.15 &1.51$\pm$0.31&1.25$\pm$0.30 \\ \hline
\end{tabular}
\caption{The mean and standard deviations of the activation ratios $s1$, 
             $s2$ and $s3$ considering the concentric levels with the 
             first and second highest number of nodes in the morphological
             networks with $n=1, 3, 5, 10$ and $20$.  $T=7$.
        }\label{tab:s1s2s3_short}
\end{table*}

Most networks involve at least one ratio with values smaller or near
$1$, suggesting that the activation in these networks proceed in a
relatively gradual manner.  Indeed, the activations tend to be more
gradual for the configurations with larger number of spokes.

\section{Characterization of the Dynamics}

\begin{figure}[htb]
  \begin{center}
  \includegraphics[width=1\linewidth]{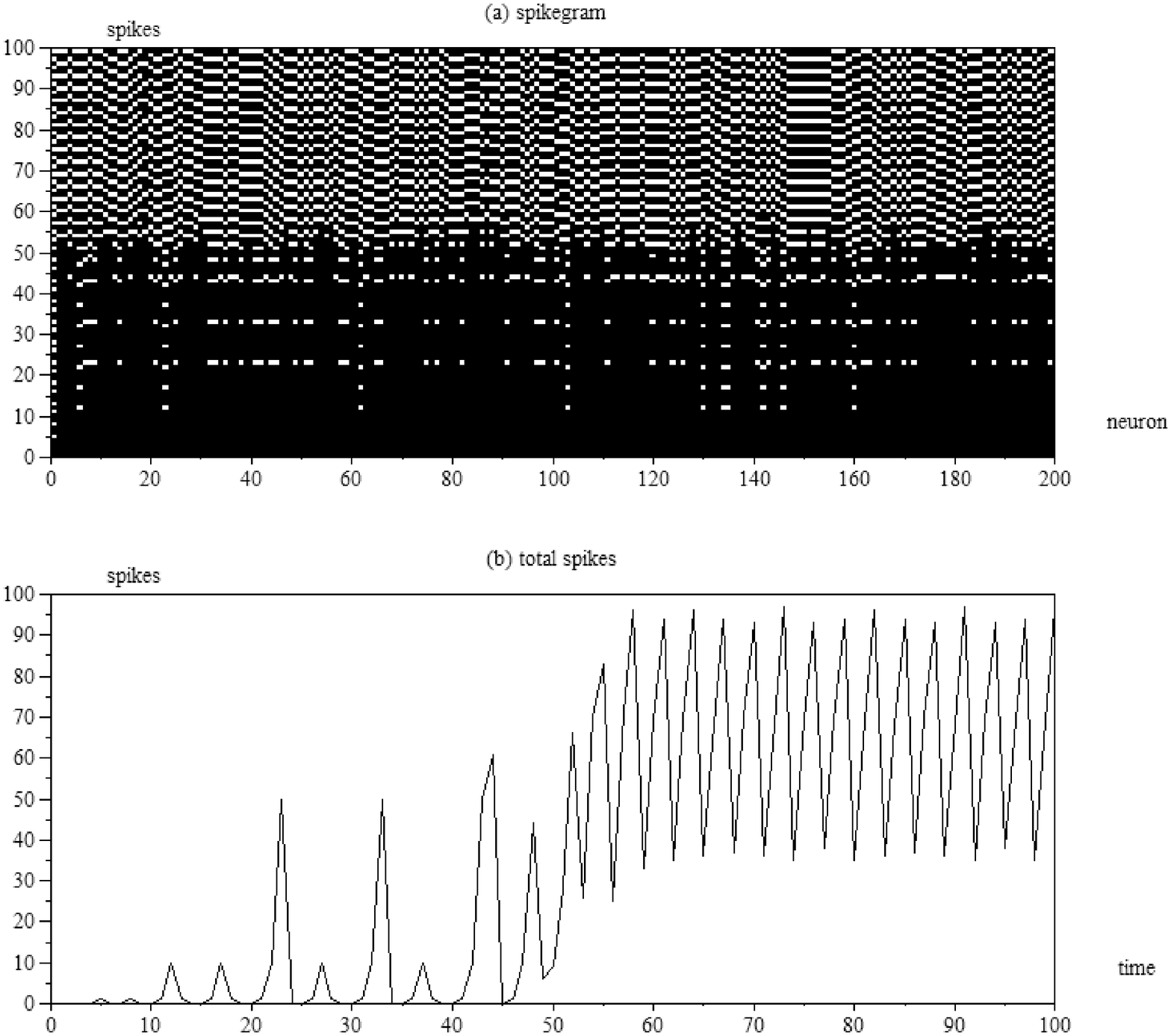} 
  \caption{The spikegram (a) and total number of spikes along time
               (b) obtained for a morphological network with 
                $N=200$, $n=3$ and $T = 3$.
  ~\label{fig:ex_dyn}}
  \end{center}
\end{figure}

In this section we investigate the transient activation dynamics of
the morphological networks with respect to several configurations.
Figure~\ref{fig:ex_dyn} illustrates the spikegram (a) and total number
of spikes (b) obtained for a network with $N=200$, $n=3$ and $T = 3$.

The spikegram in Figure~\ref{fig:ex_dyn}(a) shows the occurrence of a
spike (shown in white) over time (vertical axis) for each of the
$N=200$ neurons in the network (horizontal axis).  As the network is
fed from node 1, a few neurons (the most immediate neighbors of node
1) start sporadic firing, up to nearly $t=50$, when most of the
neurons begin producing spikes, signaling the occurrence of an
avalanche.  Figure~\ref{fig:ex_dyn}(b) shows the total number of
spikes along time.  In the case of this specific example, this signal
involves three preliminary peaks of spiking, followed by the
avalanche. The total number of spikes tends to undergo regular
oscillations after the avalanche.

\begin{figure}[htb]
  \begin{center}
  \includegraphics[width=1\linewidth]{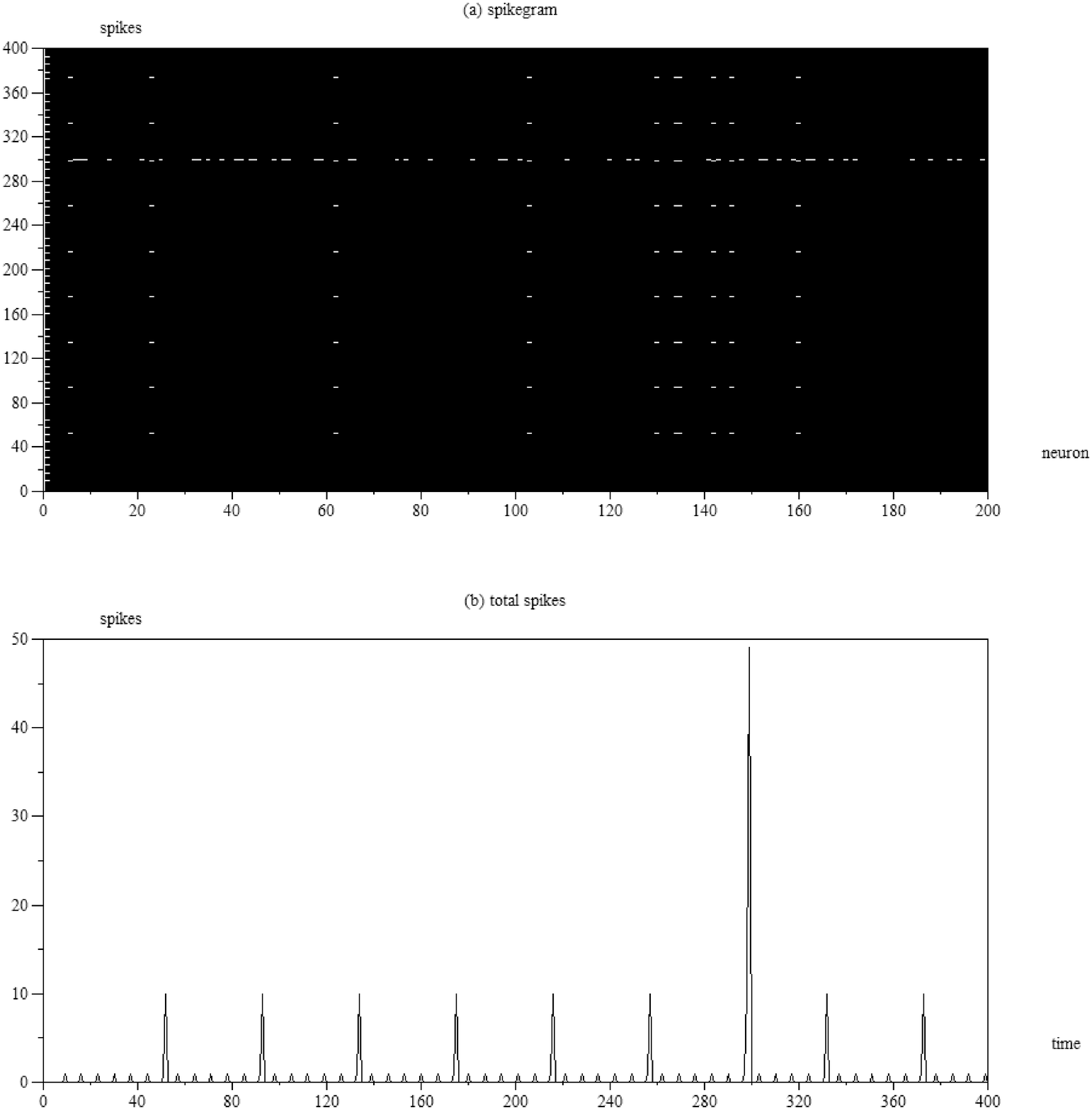} 
  \caption{The spikegram (a) and total number of spikes along time
               (b) obtained for a morphological network with 
                $N=200$, $n=3$ and $T = 7$.
  ~\label{fig:gradual}}
  \end{center}
\end{figure}

As predicted by the respective three activation ratios calculated for
each of the networks configuration (see previous Section), gradual
activations and almost complete lack of avalanches were observed for
$T=7$, as illustrated in Figure~\ref{fig:gradual}.  In other words,
the configurations of morphological networks considered in the present
work incorporate concentric organizations which imply the transient
activation to be distributed in a relatively gradual manner.

So far we have considered identical model neurons with uniformly
distributed spokes of equal length. However, real-world biological
neuronal systems are characterized by the coexistence of short and
long range connections.  The latter interconnect diverse cortical
modules and regions~\cite{Squire:2003,Zeki:1999}.  In order to
investigate the effect of long-range connections in our morphological
networks, we added a number of random edges.
Figure~\ref{fig:conc_meas} shows the three main hierarchical
measurements obtained for the same network used to produce
Figure~\ref{fig:ex_dyn}, but with $20\%E$ new randomly chosen directed
edges.  It is evident from this figure that the incorporation of
additional edges completely changes the concentric organization of
morphological networks, by reducing the total number of concentric
levels $H$ (compare with Fig.~\ref{fig:conc_meas}) as well as implying
sharper peaks in the hierarchical number of nodes and intra-ring
degrees signatures.

\begin{figure*}[htb]
  \begin{center} 
  \includegraphics[width=1\linewidth]{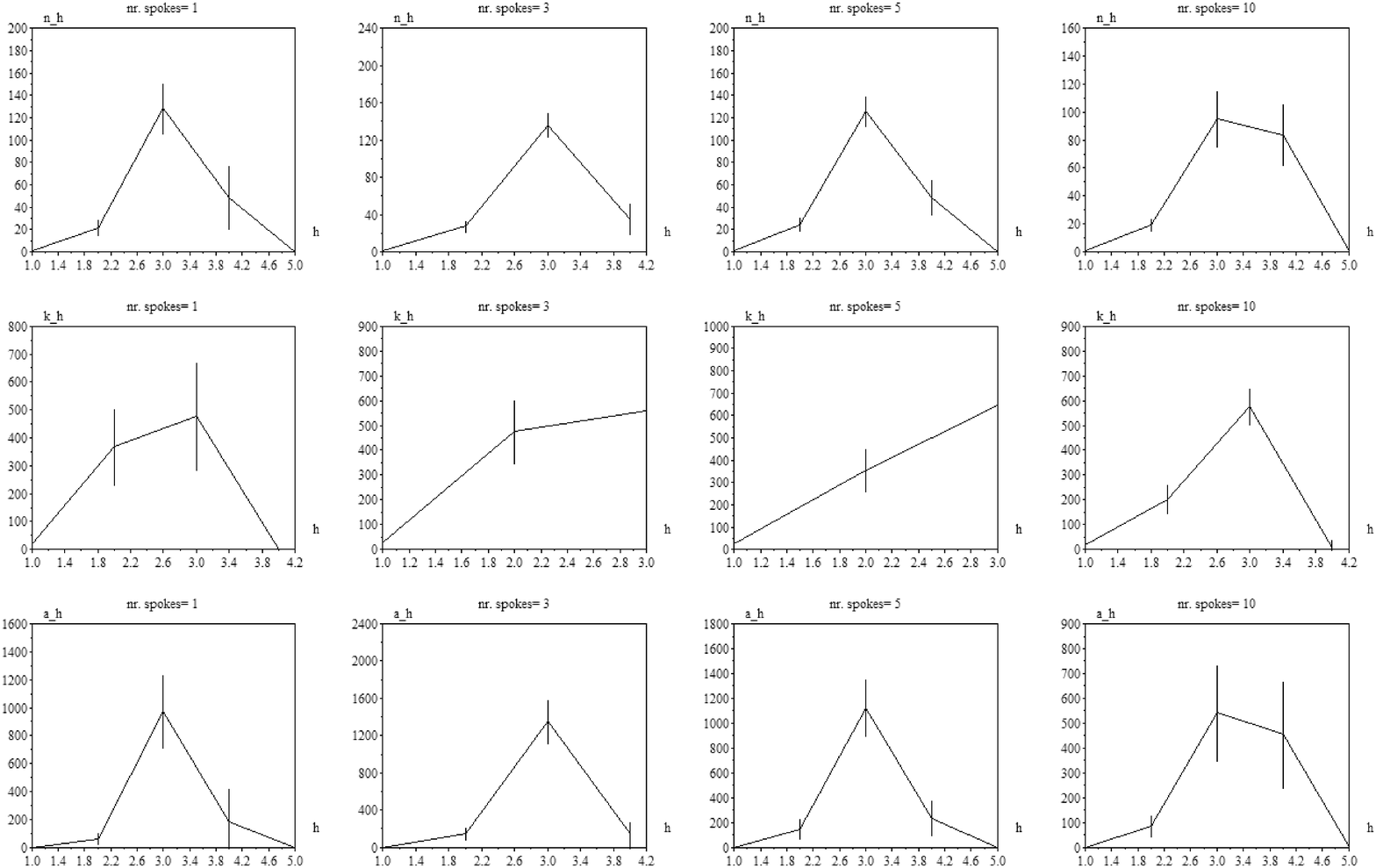}
  \caption{The hierarchical number of nodes $n_h(i)$, the hierarchical 
             degrees $k_h(i)$ and the intra-ring degrees $a_h(i)$
             obtained for the morphological networks with $N=200$
             and $n=1, 3, 5, 10$ and $20$ spokes and $20\%$ additional
             edges.
  ~\label{fig:conc_meas_add}} 
  \end{center}
\end{figure*}

Table~\ref{tab:s1s2s3_long} gives the mean and standard deviation
values of the three activation ratios obtained for the 50
morphological networks with $20\%E$ additional edges and $T=7$
respectively to the two concentric levels containing the largest
number of nodes.  It should be observed that, as a consequence of the
smaller diameter of the networks with $20\%E$ additional edges, the
level with the second highest number of nodes contains only a few
nodes. Relatively to Table~\ref{tab:s1s2s3_short}, we can see that the
addition of edges implied in substantially larger values for all the
three ratios, predicting sharper and more intense avalanches
(confirmed experimentally).  Observe that both
Tables~\ref{tab:s1s2s3_short} and~\ref{tab:s1s2s3_long} refer to
$T=7$.  Such an major change in the activation ratios is a direct
consequence of the drastic influence of the additional edges over the
concentric organization of the respective morphological networks.
Even more intense changes have been obtained for larger number of
added edges.

\begin{table*}
\centering
\begin{tabular}{|c||c|c|c|c|c|c|}  \hline  
$n$    &  $s1 (first)$   & $s1 (second)$     & $s2 (first)$    &  $s2 (second)$         & $s3 (first)$    & $s3 (second)$    \\  \hline
$n=1 $ &3.53$\pm$0.73&1.18$\pm$0.53 &2.54$\pm$0.26&1.93$\pm$0.32 &5.60$\pm$0.89&3.50$\pm$3.50 \\ \hline
$n=3 $ &5.45$\pm$0.69&1.15$\pm$0.21 &3.28$\pm$0.33&1.61$\pm$0.54 &5.77$\pm$0.73&5.49$\pm$4.65 \\ \hline
$n=5 $ &4.60$\pm$0.74&1.07$\pm$0.38 &2.96$\pm$0.35&1.66$\pm$0.50 &4.85$\pm$0.74&2.97$\pm$3.07 \\ \hline
$n=10$ &2.82$\pm$1.07&1.44$\pm$0.76 &2.06$\pm$0.23&1.49$\pm$0.28 &3.40$\pm$0.66&2.26$\pm$0.76 \\ \hline
$n=20$ &2.55$\pm$0.73&1.42$\pm$0.64 &1.61$\pm$0.19&1.17$\pm$0.27 &2.33$\pm$0.49&2.65$\pm$0.66 \\ \hline
\end{tabular}
\caption{The mean and standard deviations of the activation ratios $s1$, 
             $s2$ and $s3$ considering the concentric levels the 
             first and second highest number of nodes in the morphological
             networks with  $20\%E$ additional edges and 
             $n=1, 3, 5, 10$ and $20$.  $T = 7$.
        }\label{tab:s1s2s3_long}
\end{table*}

The above results imply that morphological networks involving short
and long-range connections are substantially more likely to exhibit
avalanches of activation than morphological networks involving only
short-range connections.  Such a finding presents important potential
implications for neuroscience, especially because it shows that the
activation of morphological networks, including their susceptibility
to avalanches, can be controlled by the density of short and
long-range links.

\section{Relating Structure and Dynamics}

Having investigated the integrate-and-fire dynamics for several
configurations of morphological networks, our attention is now focused
on the particularly relevant issue of relating the structural and
dynamical aspects of the considered networks.

Three measurements of each avalanche were extracted automatically: (i)
its onset time $t_i$; (ii) the mean of the total number of spikes
after the avalanche takes place $\left< N_s \right>$, and (iii) the
respective standard deviation $\sigma_{N_s}$.  The avalanches had to
be detected before such measurements could be calculated.  This was
achieved by thresholding the total number of spikes with signals at
one fifth of the maximum height. A total of 500 time steps was
considered.
 
\begin{figure*}[htb]
  \begin{center} 
  \includegraphics[width=1\linewidth]{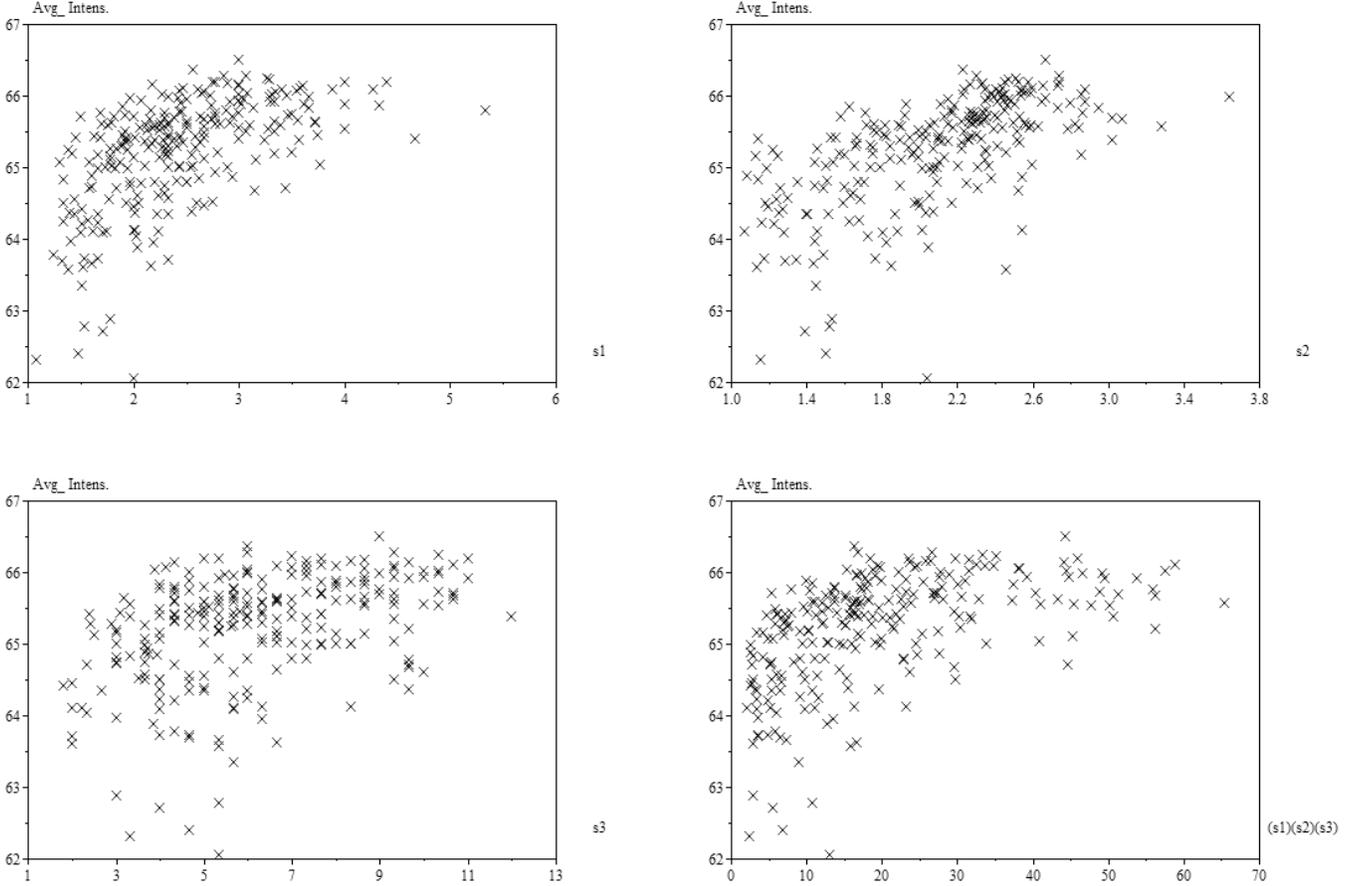}
  \caption{The scatterplots of the average total number of spikes
              along time ($N=200$ and $T=3$) and respective
              activation ratios.
  ~\label{fig:corrs}} 
  \end{center}
\end{figure*}

Figure~\ref{fig:corrs} shows the scatterplots of $\left< N_s
\right>$ against $s1$ (a), $s2$ (b), $s3$ (c), as well as the product
$(s1)(s2)(s3)$, obtained for $N=200$, and $T=3$ for all values of $n$.
This figure shows that the mean intensity of the avalanches tends to
increase steeply with the activation ratios and then saturate at a
value of about 65.  This results shows that important features of the
integrate-and-fire dynamics, such as the average intensity of the
avalanches, are intrinsically related to the respective structural
properties of the network, here represented by the three activation
ratios, which are ultimately derived from the concentric
organization. No relationships have been identified between the
activation ratios and the other two avalanche measurements.

\section{Concluding Remarks}

The intersection between complex networks and neuronal networks, which
has been called \emph{complex neuronal networks}, represents one of
the most challenging and promising research areas because it allows
neuroscience to be revisited while considering the relationship
between structured connectivities such as small-world and scale-free
and the respectively obtained dynamics.  The current article has
addressed this important paradigm with respect to more biologically
realistic networks, called \emph{morphological networks}, whose
connectivity is defined as a consequence of the shape and distribution
of geometrical neuronal cells.  By assuming a uniformly random spatial
distribution of geometrically simple neurons consisting of a circular
axon and a set of straight dendrites radiating from the soma with
uniform angles, it was possible to keep the number of involved
parameters as small as possible, allowing the systematic investigation
of the effect of the shape of the neurons on the overall topological
properties of the resulting networks as well as on the respective
integrate-and-fire dynamics.  Several interesting results were
obtained:

\emph{Characterization of the Topology of Morphological Networks in
Terms of Traditional Measurements:} The topology of the considered
morphological networks was characterized in terms of several
measurements, including average out-degree, clustering coefficient,
symmetry index, and size of the largest connected component.  We found
that networks obtained by considering neurons with larger number of
dendrites tended to exhibit greater symmetry and larger connected
components.  The average out-degree and clustering coefficient tended
to increase with the number of dendrites up to a maximum, decreasing
thereafter.

\emph{Characterization of the Topology of Morphological Networks in
Terms of Hierarchical Measurements:} The concentric organization of
the morphological networks, as revealed by the hierarchical number of
nodes, hierarchical degree and intra-ring degree, exhibited several
concentric levels, yielding respective signatures characterized by a
peak nearly the intermediate level.  The gradual distribution of these
measurements along the concentric level suggests more gradual
activation of the respective networks.

\emph{Prediction of the Type of Activation of the Networks:} The
concentric organization of a network has been found to define
important features of the respective integrate-and-fire
dynamics~\cite{Costa_equiv:2008}.  We considered the three activation
indices introduced in~\cite{Costa_lattice:2008} in order to predict
the type of activation of the network.  The higher the value of such
ratios --- which are obtained from the hierarchical number of nodes,
hierarchical degree and intra-ring degree, the higher the probability
of getting abrupt activation of the network and onset of avalanches.
Because the activation ratios for the concentric levels with the
largest number of nodes are particularly small, the morphological
networks considered in this work are expected to undergo relatively
smooth dissemination of the activation received from a source node.
Such a type of dynamics was experimentally confirmed.

\emph{Investigation of Morphological Networks Involving Short- and
Long-Range Connections:} Because biological neuronal networks are
known to incorporate short- and long-range connections, we repeated
our investigations of structure and dynamics considering morphological
networks involving also long-range dynamics.  It was found that the
incorporation of additional connections can change drastically the
concentric organization of morphological networks, implying
substantial changes also in the respective integrate-and-fire
dynamics.  In particular, it was found that the addition of $20\%$
random directed edges reduced substantially the number of concentric
levels while increasing the activation ratios.  The so-obtained
morphological networks were found to undergo sharp activation, with
onset of avalanches.  Such a finding implies that the way in which
biological networks are activated can be controlled by the ratio
between short- and long-range connections, with morphological networks
involving only short-range connections undergoing more gradual
activation.

\emph{Characterization of the Activation Dynamics in Terms of the
Average Intensity of Avalanches and Structure-Dynamics Relationships:}
A specific methodology was developed in order to automatically
identify the presence of avalanches from the total number of spikes
signatures in terms of time.  This also allowed the identification of
the initiation time and average and standard deviation of the total
number of spikes after the occurrence of the avalanches.  In order to
relate this dynamical behavior to the topology of the networks we
investigated possible relationships between the activation ratios
(derived from the concentric organization) and initiation time of
avalanches, and the average and standard deviation of the total number
of spikes after their respective onset.  It was verified that the
average of the total number of spikes after the avalanches tend to
increase steeply with any of the three activation ratios, saturating
at a maximum value.  No relationships were observed between the ratios
and the other two avalanche features.

Possible future directions of research include the consideration of
neurons with non-uniformly distributed spokes, neurons with higher
branching orders, as well as different types of neurons in the same
network and non-uniformly distributed neurons.  It would also be
interesting to consider integrate-and-fire dynamics with internal
activation decay~\cite{Costa_activ:2008}.

\vspace{1cm}
{\bf List of Symbols}

\begin{trivlist}

\item  $\Gamma =$ a graph or complex network;

\item  $N =$ total number of nodes in a network;

\item  $E =$ total number of edges in a network;

\item  $K =$ the adjacency matrix of a complex network;

\item  $k(i) =$ degree of a network node $i$;

\item  $ik(i) =$ in-degree of a network node $i$;

\item  $ok(i) =$ out-degree of a network node $i$;

\item  $cc(i) =$ clustering coefficient of a network node $i$;

\item  $C = $ the size of the largest strongly connected component
                      in the network;

\item  $L = $ the size of the workspace used to build the morphological
              networks (i.e. $L \times L$);

\item  $n(i) = $ the number of spokes (dendrites) of a neuron $i$;

\item  $r(i) = $ the radius of the soma (also axon);

\item  $S(i) = $ the current internal activation of neuron $i$;

\item  $T(i) = $ the threshold of neuron $i$;

\item  $u(i) = $ the number of incoming connections at neuron $i$;

\item  $v(i) = $ the number of outgoing connections of neuron $i$;

\item  $H = $ the total number of concentric levels in a network;

\item  $n_h(i) = $ the number of hierarchical nodes at level $h$ of
                     neuron $i$;

\item  $k_h(i) = $ the hierarchical degree at level $h$ of neuron $i$;

\item  $a_h(i) = $ the intra-ring degree at level $h$ of neuron $i$;

\item  $s1_h(i) = $ the forward activation ratio at level $h$ 
                       of neuron $i$;

\item  $s2_h(i) = $ the reflexive activation ratio at level $h$ 
                       of neuron $i$;

\item  $s3_h(i) = $ the backward activation ratio at level $h$ 
                       of neuron $i$.

\end{trivlist}

\begin{acknowledgments}
Luciano da F. Costa thanks CNPq (301303/2006-1) and FAPESP (05/00587-5)
for sponsorship.
\end{acknowledgments}

\bibliography{morph_nets}

\end{document}